\documentclass[aps,prd,twocolumn,groupedaddress, showpacs]{revtex4}
\usepackage{graphics}
\usepackage{color, amsmath, epsfig, amssymb,mathrsfs}

\begin{document}

\preprint{Amitai's paper}

\title{Relativistic Images in Randall-Sundrum II Braneworld Lensing}

\author{Amitai Y. Bin-Nun}
\email[]{binnun@sas.upenn.edu}
%\homepage[]{Your web page}
%\thanks{what}
%\altaffiliation{}
\affiliation{Department of Physics and Astronomy, University of Pennsylvania}

\date{\today}

\begin{abstract}
In this paper, we explore the properties of gravitational lensing by black holes in the Randall-Sundrum II braneworld. We use numerical techniques to calculate lensing observables using the Tidal Reissner-Nordstrom (TRN) and Garriga-Tanaka metrics to examine supermassive black holes and primordial black holes. We introduce a new way tp parameterize tidal charge in the TRN metric which results in a large increase in image magnifications for braneworld primordial black holes compared to their 4 dimensional analogues. Finally, we offer a mathematical analysis that allows us to analyze the validity of the logarithmic approximation of the bending angle for any static, spherically symmetric metric. We apply this to the TRN metric and show that it is valid for any amount of tidal charge. 
\end{abstract}


\maketitle
\section{\label{sec:intro} Introduction}

Gravitational lensing provided the first experimental verification of General Relativity (GR) through observations of starlight bending around the Sun during an eclipse in 1919 \cite{schneider, 1919eclipse}. Gravitational lensing is a very important probe of cosmological and astrophysical questions \cite{lensingreview, jainreview}. The vast majority of lensing studies are in the weak deflection limit, with light rays bent on the order of, at most, arcseconds. Over the last decade, there has been a renewed interest in strong deflection limit lensing and using it as a probe of GR. Some studies use a numerical method \cite{VE2000, VE2002, V2009, VK2008} and a more analytical approach has been used as well \cite{bozzaetal2001, eiroad, bozza2002, bozzas2, bozzas2revised, bozzamancini2009, eiroarn}.  In the last few decades, the mass of the supermassive black hole (SMBH) at the center of the galaxy (Sgr A$^\ast$) has become well known \cite{smbhdetails, propSgrA}, so the galactic center is often cited in the above lensing studies as an ideal candidate for the study of strong deflection limit lensing \cite{VE2000, VE2002, V2009, VK2008, bozzas2, bozzas2revised, bozzamancini2005, whisker2005}. 

As the trajectory of a photon brings it closer a black hole's photon sphere, it undergoes a growing bending angle, and if the point of closest approach is incident with the photon sphere, this bending angle goes to infinity. This gives rise to a theoretically infinite sequence of images close to the photon sphere on both sides of the optic axis due to photons looping around the black hole before reaching the observer. These images are termed relativistic images and their use as a probe of GR was pioneered by \cite{VE2000} and expanded to include the topic of naked singularities and cosmic censorship in \cite{VE2002, VK2008}. However, as noted in these articles, relativistic images are highly demagnified and their observational prospects are very dim at present. A later analysis \cite{bozzas2} showed that a relatively ``bright" relativistic image (brighter than the 32$^{\text{nd}}$ magnitude in the K-band) may be visible in late 2017 through early 2018. Further investigations by \cite{bozzas2revised, bozzamancini2009, depaolis} discuss prospects for observation of secondary images that undergo a large bending angle. Studying the properties of these secondary images requires some of the techniques coming from studies in the strong deflection limit. As observation of images lensed in the strong deflection limit is a distinct possibility, a close study of their properties is warranted.
As mentioned above, \cite{bozzaetal2001, bozza2002, bozza2007} introduce a formalism that simplifes the calculation of observables for lensing in the strong deflection limit and introduces simple formulae for image positions and magnifications. This formalism can be used for any spherically symmetric metric and \cite{whisker2005, mmreview, eiroad, aliev} expand the analysis of strong field lensing to metrics which come from the Randall-Sundrum (RS) II braneworld scenario, a theory that has evoked great interest in the last decade.
In Sec. \ref{sec:sfl}, we review the two methodologies for solving the lens equation for lensing observables in the strong deflection limit, the numerical method of \cite{VE2000}, and the analytical method based on approximating the bending angle with a logarithmic term by \cite{bozza2002, bozza2007}. In Sec. \ref{sec:bbh}, we introduce the Randall-Sundrum braneworld and discuss several 4 dimensional solutions on the brane for the 5 dimensional black hole of RS II theory. We comment on several metrics that come from this theory and their regime of applicability. In Sec. \ref{sec:bbhl}, we calculate the observational properties of a galactic object lensed by the supermassive black hole at Sgr A*, and a galactic source lensed by a primordial black hole in our solar system. We compare the observables when modelling the black hole with the Schwarzschild metric as well as with several braneworld black hole metrics. In Sec. \ref{sec:ana}, we introduce a general test for the logarithmic approximation of the bending angle for all spherically symmetric metrics and show that it works for all values of tidal charge.  Sec. \ref{sec:conc} contains a discussion of the results and possible future research directions.

\section{\label{sec:sfl} Strong Deflection Limit Lensing} 

\begin{figure}[h]
 \begin{center}
 \includegraphics[width=0.4 \textwidth]{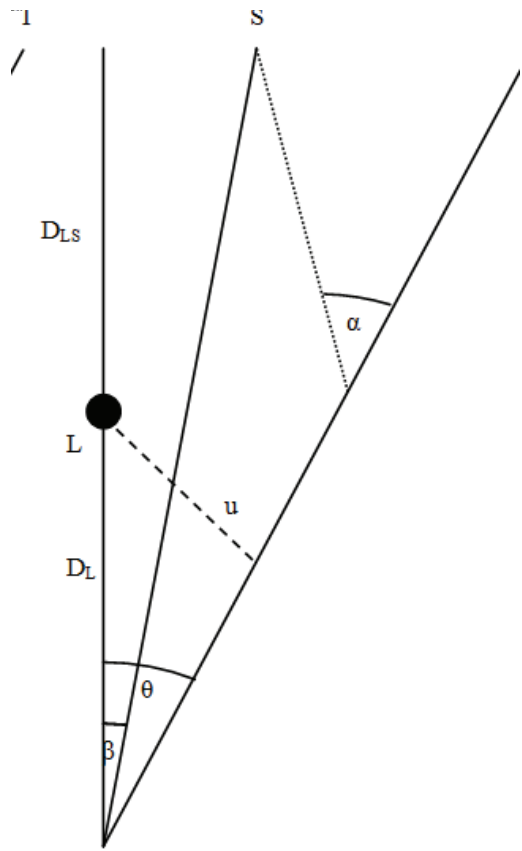}
 \caption{The simplest system used to study gravitational lensing. A source at $S$, a lens at $L$, and an observer.}
 \label{fig:lensdia}
 \end{center}
 \end{figure}

 In Sec. \ref{sec:intro}, we discussed the importance of lensing in the strong deflection limit. To go beyond the weak deflection limit for the bending angle, we will need to derive lensing quantities directly from the metric. In this paper, we will only be considering metrics of the spherically symmetric form:

\begin{equation}
ds^2= -A(r)dt^2 + B(r) dr^2 +C(r) d\Omega^2
\label{metric}
\end{equation}
The weak deflection limit approximation is only valid in the limit of a small bending angle and when the point of closest approach is far from the black hole. When the point of closest approach is close to the black hole, the bending angle is derived from the equations of motion  \cite{weinberg}

\begin{equation}
g_{\mu \nu} \dot{x}^{\mu}\dot{x}^{\nu}=0
\label{eq:eom}
\end{equation}
to be
\begin{eqnarray}
\nonumber \alpha (r_0) &=& 2 {\int_{r_0}}^{\infty}\left(\frac{B(r)}{C(r)}\right)^{1/2}
                       \left[(\frac{r}{r_0})^2\frac{C(r)}{C(r_0)}\frac{A(r_0)}{A(r)}-1\right]^{-1/2} \\ 
& \times & \frac{dr}{r}- \pi
      \label{bending}
\end{eqnarray}
where $r_0$ is the point of closest approach. Once we can calculate the bending angle of a null geodesic as a function of $r_0$, the results are used in conjunction with the Virbhadra-Ellis lens equation \cite{VE2000}:

\begin{equation}
  \tan\beta =  \tan\theta -  \frac{D_{LS}}{D_S} [ \tan \theta + \tan(\alpha - \theta)]
  \label{GravLensEqn}
\end{equation}
where $D_{LS}$, $D_L$, $\theta$, and $\beta$ are, respectively, the distance from the lens plane to the source plane, the distance from the observer to the lens plane, the angle of the image relative to the optic axis, and the angle of the source relative to the optic axis. $D_S$, which is not pictured in  Fig. \ref{fig:lensdia} is the total distance from the observer plane to the source plane and is just $D_{LS} + D_L$. In \cite{bozzalens}, there is a discussion of lens equations used in gravitational lensing. While concluding that the ``improved Ohanian" lens equation \cite{ohanian, bozzalens} is more accurate than Eq. (\ref{GravLensEqn}), \cite{bozzalens} notes that the Virbhadra-Ellis lens equation is precise to within a factor of $10^{-4}$ in most situations. This suffices for the purposes of this paper and offers the best way to compare our results with existent work, the majority of which is done with the Virbhadra-Ellis equation. While this is not an exact lensing solution \cite{frittellietal1999}, this approximation is accurate and easy to use for calculating relativistic images \cite{bozzalens}. A typical gravitational lensing scenario is pictured in Fig. \ref{fig:lensdia}.

In a spherically symmetric scenario, the magnification of an image is given by 

\begin{equation}             
  \mu = ( \frac{\sin{\beta}}{\sin{\theta}} \ \frac{d\beta}{d\theta} )^{-1} 
  \label{Mu}
\end{equation}
There are two distinct approaches in going from the bending angle to the observables of image positions and image magnifications. First we will review the approach of \cite{bozzaetal, bozza2002, bozza2007} that approximates the bending angle with a logarithmic term. The results in this section can be applied to any spherically symmetric, static metric. Then we will review the completely numerical approach of \cite{VE2000, VE2002, VK2008,V2009} in a Schwarzschild spacetime. This paper extends the numerical results in the literature to braneworld black holes. In a later section of the paper (Sec. \ref{sec:ana}), we will show a general method for showing that the analytical approach reproduces the numerical approach for a particular spacetime.

\subsection{Analytical Solution}

While an exact solution to the bending angle without reference to a background metric is given by \cite{frittellietal1999, frittellietal2000} with solutions being given in integral form, the earliest attempt at trying to develop an analytic approach to relativistic images was done by \cite{bozzaetal2001}, where they attempted to express Eq. (\ref{bending}) as an expansion of an elliptic integral. They then found the first order expansion of the elliptic equation from its divergence at the photon sphere.  This approach is later used to calculate lensing observables in the braneworld scenario \cite{eiroad}. The approach we will be working with in this paper was developed by \cite{bozza2007} and we will develop the formalism with the intent of demonstrating its range of effectiveness in Sec. \ref{sec:ana}.

The equations of motions in Eq.(\ref {eq:eom}) have cyclic coordinates $t$ and $\phi$ leading to conserved quantities
\begin{eqnarray}
E &=& A(r) \dot{t}\\
J &=& C(r) \dot{\phi}
\end{eqnarray}
and the following expression for $\dot{r}$:
\begin{equation}
\dot{r}= \pm \frac{E}{\sqrt{BC}}\sqrt{\frac{C}{A}-\frac{J^2}{E^2}}
\label{eq:rdot}
\end{equation}
where an overdot sepresents a derivative with respect to the affine parameter. The radial and angular motion of the null geodesic can then be characterized by the ratio of the two conserved quantities:
\begin{equation}
u \equiv\frac{J}{E}
\end{equation}
In order for a photon which is initially travelling with $\dot{r}<0$, Eq. (\ref{eq:rdot}) must vanish for the photon to invert its motion and not fall into the black hole. For a given value of $u$ and assuming the function $\frac{C}{A}$ has a single minimum, $\dot{r}=0$ will occur when 
\begin{equation}
\frac{C(r_0)}{A(r_0)}=u^2
\end{equation}
where $r_0$ is the point of closest approach. Since  $\frac{C}{A}$ is lower-bounded, this equality can only be satsfied if $u$ is greater than the minimum value
\begin{equation}
u_m= \sqrt{\frac{C_m}{A_m}}
\end{equation}
 where we have defined $A_m \equiv A(r_m)$. The point $r_m$ represents the radial coordinate of the photon sphere. 

These terms suggest the rewriting of Eq. (\ref{bending}) as 
\begin{equation}
\Delta \phi_i={\int_{r_0}}^{D_{i}}  u \sqrt{\frac{B}{C}}(\frac{C}{A}-u^2)^{-1/2} dr
\label{eq:int}
\end{equation}
with the notation $D_{i}= D_{LS}, D_L$, breaking the bending angle integral into two parts- the first for the infall and the second for the portion of the photon's path from the radial minimum to the observer. To perform a detailed analysis of the bending angle around $r_0=r_m$, we will examine the function
\begin{equation}
R(r,u)= \frac{C(r)}{A(r)}-u^2
\end{equation}
From the previous discussion, we have shown that $R(r,u)$ has a minimum at $r_m$ for any $u$. It also vanishes at $(r_0, u)$ by definition of $r_0$ and at $(r_m, u_m)$ by definition of $u_m$. We are interested in the properties of the bending angle corresponding to an inversion point very close to the photon sphere $r_m$, so we parameterize the inversion point as
\begin{equation}
r_0=r_m (1+ \delta)
\end{equation}
which corresponds to an impact parameter that is also very close to the minimum
\begin{equation}
u=u_m (1+ \epsilon)
\end{equation}
These properties allow for expansion of $R(r,u)$  in orders of $\epsilon$ and $\delta$. The lowest order non-vanishing terms are:
\begin{equation}
R(r,u)=0= \frac{1}{2}\frac{\partial^2 R}{\partial r^2}(r_m,u_m)r_m^2 \delta^2+ \frac{\partial R}{\partial u}(r_m,u_m)u_m \epsilon
\label{eq:firstorder}
\end{equation}
and this creates the simple relationship between $\epsilon$ and $\delta$ of
\begin{equation}
\epsilon= -\frac{\beta_m}{2 u_m^2}\delta^2
\label{eq:relation}
\end{equation}
where 
\begin{equation}
\beta_m= \frac{1}{2}\frac{\partial^2 R}{\partial r^2}(r_m,u_m)
\end{equation}

We have gone through this part of the derivation in detail because we will comment in Sec. \ref{sec:ana} about the validity of leaving off the higher-order terms in $\epsilon$ and $\delta$. From here, \cite{bozza2007} uses this expansion of $R(r,u)$ to perform the integration in Eq. (\ref{eq:int}). Continuing along these lines, they define  the position of the source ($D_{LS}$) and the observer ($D_L$) with the coordinate $\eta$ defined by:
\begin{equation}
r=\frac{r_0}{1-\eta}
\end{equation}
with $0 \leq \eta \leq 1$.  Eventually \cite{bozza2007} finds that the bending angle is:
\begin{equation}
\Delta \phi= a \text{log} \frac{4 \eta_O \eta_S}{\delta^2}+ b_O +b_S
\label{eq:bendlog}
\end{equation}
with the subscript $O$ and $S$ corresponding to the observer and source, respectively, and $a$ and $b_i$ are functions of the metric
\begin{eqnarray}
a&=&r_m \sqrt{\frac{B_m}{A_m \beta_m}}\\
b_i&=& \int_{0}^{\eta_i} d \eta \: Sign (\eta) g(\eta) \\
\nonumber g(\eta)&=&  u_m  \sqrt{\frac{B(\eta)}{C(\eta)}}[R(\eta,u_m)]^{-1/2} \frac{r_m}{(1-\eta)^2}- \\
&& \frac{u_m}{\sqrt{\beta_m}}\sqrt{\frac{B_m}{C_m}}\frac{r_m}{|\eta|}
\end{eqnarray}
This result is equivalent to the result in \cite{bozza2002} when $\eta_O=\eta_S=1$. In order to calculate image positions and mangifications, we use the formalism of \cite{bozza2002}
\begin{eqnarray}
&& \overline{a}= \frac{a}{2}\\ 
&& \overline{b}=-\pi + \int_{0}^{1} g(\eta) d \eta + \overline{a} log \frac{2 \beta_m}{A_m}
\end{eqnarray}
and adopt Eq. (9) from \cite{bozza2002}
\begin{equation}
\alpha(\theta)=\overline{a} \text{log} (\frac{\theta D_L}{u_m}-1)+\overline{b}
\end{equation}
Using this logarithmic term in conjunction with a simplified lens equation
\begin{equation}
\beta= \theta-\frac{D_{LS}}{D_S}\Delta \alpha_n
\end{equation}
where $\Delta \alpha_n$ is the bending angle once $2 \pi n$ has been subtracted to account for the ``loops" that the photon has made, we can easily solve for $\theta_0^n$ which gives us the image position that corresponds to a bending angle $\alpha(\theta)= 2 \pi n$:

\begin{eqnarray}
&& \theta_n^0=\frac{u_m}{D_L} \left(1+e_n \right) \label{theta0}\\
&& e_n=e^{\frac{\overline{b}-2n\pi}{\overline{a}}} \label{en}
\end{eqnarray}
The image position for a given source position $\beta$ is then given by $\theta_0^n$ and a correction term:
 \begin{equation}
\theta_n = \theta_n^0+\frac{ u_m e_n\left(\beta-\theta_n^0
\right) D_{OS}}{\overline{a} D_{LS}D_L} .
\label{Images}
\end{equation}
where the correction is much smaller than $\theta_0^n$. It is straightforward to derive the magnification of the $n^{\text{th}}$ relativistic image as:
\begin{equation}
\mu_n =e_n \frac{ u_m^2\left(1+e_n \right)
D_{OS}}{\overline{a} \beta D_L^2 D_{LS}} 
\label{Magnification}
\end{equation}
This outlines the analytical approximation in the strong-deflection limit. We will use it in this paper to briefly compare results with the purely numerical method. Also, we will examine if there are circumstances under which the higher order terms in Eq. (\ref{eq:firstorder}) would be relevant and therefore interfere with this form of the strong deflection limit approximation. We now review the numerical algorithm for gravitational lensing.

\subsection{Numerical Solution}

In the numerical procedure first outlined by \cite{VE2000, VE2002}, conservation of $J$ and $E$  is used to relate the image position with the point of closest approach. All the equations in this section up to here are true of any spherically symmetric, static metric. However, since numerical techniques will be employed in this section, the functions of the metric must be specified, and this section assumes the use of a Schwarzschild metric. In addition, this section and the rest of this paper uses geometric units in which $c=G=1$ \cite{schneider}. This approach can be applied to any metric, as it is in Sec. \ref{sec:bbhl}. The impact parameter $u$ is conserved, and in the asymptotic Minkowski space (which is required for use of Eq. (\ref{GravLensEqn})), the perpendicular distance from the center of the lens to the null geodesic is

\begin{equation}
u=D_L \: \textrm{Sin} \: \theta
\label{impact}
\end{equation}
This is only valid if the source is in the asymptotic region ($D_{LS} >> 2M$). At the point of closest approach, solving the geodesic equations yields

\begin{equation}
u=r_0 (1- \frac{2M}{r_0})^{-\frac{1}{2}}
\label{closeapproach}
\end{equation}
To simplify, we use the following definitions:

\begin{eqnarray}
 x & \equiv & \frac{r}{2M} \\
 x_0 & \equiv & \frac{r_0}{2M}
\label{r2mx}
\end{eqnarray}
to make the coordinates in terms of ``Schwarzschild radii."

Since Eq. (\ref{impact}) is equivalent to Eq. (\ref{closeapproach}),  we can now describe the image position in terms of the coordinate of closest approach (as well as converting back from $x_0$ to $\theta$):

\begin{equation}
\textrm{Sin} \: \theta=  \frac{2M}{D_L} \frac{x_0}{\sqrt{1-\frac{1}{x_0}}}
\end{equation}

For a given value of $\beta$ we insert Eq. (\ref{bending}) into Eq. (\ref{GravLensEqn}) and solve for $x_0$. This gives us a solution of the lens equation. Since Eq. (\ref{GravLensEqn}) is transcedental, there are infinite positive and negative solutions corresponding to images on both sides of the lens. The solution with the highest (lowest) value of $x_0$ represents the classic lensing solution for the primary (secondary) image. To find a relativistic image, we must determine what range the bending angle would be in to yield a relativistic image. While it is possible that a source will align nearly perfectly with the optic axis and have a smaller angular position than the relativistic image \cite{V2009}, in reality, a relativistic image will be closer to the optic axis (the line connecting the observer and source planes perpindicualar to both planes) than the source. In \cite{bozzamancini2009, bozzas2}, realistic possibilities for strong field images in the center of our galaxy do not occur with stars that are highly aligned with the optic axis (as the orbit of known stars near the center of the galaxy do not pass close to the optic axis). Hence, relativistic images will form at a smaller angle from the optic axis then the source angle and the bending angle for the first relativistic image on the side of the lens is slightly less than $2 \pi$. Therefore, to find the correct bending angle that solves Eq. (\ref{GravLensEqn}), the search for a correct value $x_0$ should be in the region that yields a bending angle of $2 \pi$. So, the first step is to compile a table of the values of $x_0$ that yield the values of $n \pi$. This is found in Table \ref{bendingx0}. To find a specific relativistic image, be it higher order on either side of the lens, the bending angle must be estimated (it will be close to $2 \pi n$ for $n^{th}$ order images with sources close to the optic axis) and corresponding values for $x_0$ searched near the appropriate value.

\begin{table}[t]

\begin{tabular}{c|c c c}
\hline
\hline
$\alpha(x_0)$ & ${x_0}_{Num}$ & ${x_0}_{Ana.}$ & $\Delta$\\
\hline
$\pi$ & 1.7603 & 1.7084 & .02947 \\
$2 \pi$ & 1.5451 & 1.5433 & .00115 \\
$3 \pi$ & 1.5091 & 1.5090 & .00005 \\
$4 \pi$ & 1.5019 & 1.5019 & $2 \times 10^{-6}$ \\
$5 \pi$ & 1.5004 & 1.5004 & $8 \times 10^{-8}$\\
\hline
\hline
\end{tabular}

\caption{\label{bendingx0} The coordinate of closest approach required to yield a particular bending angle in both the numerical technique and the analytical approximation for a Schwarzschild spacetime. The analytical approximation tends to be converge with the numerical technique as $x_0$ approaches the photon sphere, as the analytic approach loses validity as one moves away from the photon sphere. $\Delta$ is the fractional difference between the analytical and numerical result. See Sec. \ref{sec:ana} for elaboration on the validity of the analytical approach.}
\end{table}

In addition to image position, image magnification is an important observational quantity. The general formula is given by Eq. (\ref{Mu}), but incorporating Eq. (\ref{GravLensEqn}) and expressed in terms of known functions of $x_0$, magnification is given by:

\begin{eqnarray}
\nonumber \mu &=& \frac{\textrm{Sec}^2 \theta - \frac{D_{LS}}{D_S}[\textrm{Sec}^2 \theta + \textrm{Sec}^2(\alpha(x_0)-\theta)]}{\textrm{Sec}^2 \beta}\\ 
&&\times (\frac{d \alpha}{d \theta}-1)
\label{magnification}
\end{eqnarray}

Once we have solved for image position and $x_0$, the only unknown is $\frac{d \alpha}{d \theta}$ which can be broken up into

\begin{equation}
\frac{d {\alpha}}{d \theta}= \frac{d \alpha}{d x_0} \frac{d x_0}{d \theta}
\end{equation}
and the calculation of these two quantities for the Schwarzschild metric is given in \cite{VE2000}. We have just illustrated how to go from a source position to an image position (given either as an angle, $\theta$ or as a point of closest approach, $x_0$) and an image magnitude.

One of the reasons why the numerical method is carried out is to make full use of Eq. (\ref{GravLensEqn}) as the bending angle can be large and the weak field approximation does not hold. However, as Bozza points out, for a relativistic image, $\alpha$ is usually only slightly different than $2 \pi n$ and the small angle approximation can be utilized to make solving the equation more tractable, a point incorporated by Virbhadra into the numerical algorithm in \cite{V2009}. One of the advantages of the numerical method is that it is universally valid and does not depend on approximations that are only valid in the strong or weak deflection limits, an important point when studying realistic scenarios for strong deflection limit lensing \cite{bozzas2revised, bozzamancini2009}. 

\section{\label{sec:bbh} Black Holes in the Braneworld}

An important and still unverified idea in modern physics is the existence of extra dimensions. This idea had its origins in the attempts of Kaluza and Klein to unify the electromagnetic force with GR by introducing an extra dimension, which was compactified in Klein's model (See reviews in \cite{whiskerthesis, MaartensReview}). More recently, a class of these models was proposed where the Standard Model fields are constrained to a 3-brane while gravity propagates in a higher-dimensional bulk. Models which incorporate ideas along this vein are colloquially referred to as braneworld models, starting with the outbreaking \cite{add}. This paper will be based on a particular braneworld scenario of Randall and Sundrum \cite{rs1, rs2}. 

The Randall-Sundrum I model \cite{rs1} emerges from 11 dimensional M-theory when the gauge fields of the standard model are confined on two 1+9 branes located on the end points of an $S^1/Z_2$ orbifold, a scenario whose importance is highlighted by \cite{witten1} and further developed by \cite{ovrut1,ovrut2, ovrut3} . 6 dimensions are compactified, making gravity effectively 5 dimensional. The Randall-Sundrum II scenario is this model with the second brane taken to infinity \cite{MaartensReview}. The 5 dimensional metric in the absence of matter is known to be \cite{rs2}:

\begin{equation}
ds^2= e^{-2k |z|}[dt^2 - d^3 x] + dz^2
\end{equation}
where $k$ is the energy scale of the 5 dimensional cosmological constant
\begin{equation}
\Lambda_5 = -\frac{6}{l^2}=-6 k^2
\end{equation}
and $l$ is the characteristic size of the extra dimension. This cosmological constant prevents gravity from ``leaking" into the bulk at low energies.
The metric in the presence of matter is more complicated. Firstly, it is not clear whether a static black hole solution exists for a RS II braneworld. Although \cite{emparan} has shown the existence  of a static black hole in a 2+1 brane setup, it is unclear what the solution is and whether there even is a static black hole solution in the corresponding 3 + 1 brane scenario \cite{kanti, creek, shiromizu2, wiseman, wiseman2}. One approach is to write down the induced Einstein equation on the brane and attempt to solve it. Following the derivation of \cite{shiromizu}, the induced Einstein equations on the brane will be
\begin{equation}
G_{\mu \nu}= - \Lambda g_{\mu \nu}+ \kappa^2 T_{\mu \nu}+ 6 \frac{\kappa^2}{\lambda}S_{\mu \nu} -\xi_{\mu \nu}+ 4 \frac{\kappa^2}{\lambda}\mathscr{F}_{\mu \nu}
\label{eq:induced}
\end{equation}
where $\Lambda$ is the 4 dimensional cosmological constant which comes from a combination of the 5 dimensional cosmological constant and the brane tension, $\kappa$ is related to the coupling constant in 5 dimensional as well as the brane tension $\lambda$, $S_{\mu \nu}$ is a term that is second-order in the stress-energy tensor,  $\mathscr{F}_{\mu \nu}$ expresses contributions from the 5 dimensional stress-energy tensor aside from the 5 dimensional cosmological constant \cite{shiromizu, MaartensReview}, and  $\xi_{\mu \nu}$ is the double contraction of the Weyl tensor with the unit normal to the brane. The derivation of Eq. (\ref{eq:induced}) is performed in greater detail in \cite{MaartensReview}. To find the vacuum solution around a source, the following assumptions are made: The brane tension is finely tuned so that $\Lambda$ in 4 dimensional is zero. Because we deal with a vacuum solution in 4 dimensional, the terms $T_{\mu \nu}$, $\mathscr{F}_{\mu \nu}$, and $S_{\mu \nu}$ are therefore zero outside the source. This reduces the vacuum Einstein equations to

\begin{equation}
R_{\mu \nu}= -\xi_{\mu \nu}
\end{equation}
In order to find a solution to the equations, some assumptions must be made about the form of $\xi_{\mu \nu}$. A weak-field, linear perturbative expansion is known \cite{gtmetric}, but the solution has the wrong form in the strong field \cite{whiskerthesis}.  The non-linear strong field regime requires a different metric- perhaps one metric will not be sufficient to cover the entire brane spacetime, but different metrics will apply in different regimes. Attempts to study the problem have yielded several possible black hole metrics \cite{dmptidalrn, whisker2005, blackstring, casadio, gtmetric} and the effect of these metrics on lensing has been studied as well \cite{eiroad, whisker2005, petterskeeton, mmreview}. This paper will add to the literature by performing numerical studies of lensing using these metrics.  

The simplest solution, the ``black string" was found by \cite{blackstring} which desribes an effective Schwarzschild solution on the brane (by setting the Weyl term $\xi_{\mu \nu}=0$). There are a number of instabilities and difficulties with this solution \cite{gregorylaflamme}, and it is uninteresting from our perspective because the induced metric in 4 dimensions is Schwarzschild and observables in the Schwarzschild solution have already been calculated numerically. We will now set forth 3 different metrics to be studied.

\subsection{Garriga-Tanaka}

An important solution for gravity in the braneworld scenario is the Garriga-Tanaka solution \cite{gtmetric}. This metric comes from the fact that the linearized metric can be found at low energies (corresponding to scale $r >> l$). The perturbation on the brane is written in terms of a Green's function which is dominated by the low-energy zero mode of the 5 dimensional graviton. When calculating the linearized correction to the metric for a point source, the Newtonian potential, to a first-order correction is:

\begin{equation}
\frac{1}{2}h_{00}=V(r) \approx \frac{G M}{r} (1+ \frac{ 2 l^2}{3r^2})
\label{weakfield2}
\end{equation}
where $h_{00}$ is a convention used for the difference from a Minkowski spacetime of the weak field metric term $g_{00}$. Incorporating the weak field potential from Eq. (\ref{weakfield2}) yields the following metric:

\begin{eqnarray}
\nonumber dS_4^2 &=& -(1-\frac{2M}{r} - \frac{4 M l^2}{3 r^3})dt^2 \\ && +  (1+\frac{2M}{r} + \frac{2 M l^2}{3 r^3})dr^2
 + r^2 d \Omega^2
\end{eqnarray}

This is the working metric for the weak field limit, or on scales $ r >> l$. Sometimes, these limits are not identical and this will be discussed in the next section. This metric is only expected to work in the linear regime, as the Ricci scalar does not vanish as it should (as $\xi_{\mu \nu}$ is traceless) beyond first order in $M$ \cite{whiskerthesis}.

\subsection{Myers-Perry}

For small enough distance scales, we can consider the black hole as a five dimensional Schwarzschild black hole. On this length scale, the AdS curvature does not greatly affect the geometry of the black hole. Finding the metric can be done using using the Myers-Perry general form for the 4 dimensional induced metric for a higher dimensional black hole \cite{myersperry,myersperry2}. This metric is \cite{mmreview} 

\begin{equation}
dS_4^2= -(1- \frac{r_H^2}{r^2}) dt^2 + (1- \frac{r_H^2}{r^2})^{-1} dr^2 +r^2 d\Omega^2
\label{eq:metricrmp}
\end{equation}
where the black hole horizon radius $r_H$ is given by

\begin{equation}
r_H= \sqrt{\frac{8}{3 \pi}} (\frac{l}{l_P})^{\frac{1}{2}} (\frac{M}{M_P})^{\frac{1}{2}} l_P
\end{equation}

The approximate distance scale for which this metric can be considered accurate is $r << l$ because that is the domain in which the graviton does not differentiate between the AdS dimension and the other ones. $l_P$ and $M_P$ are the known standard 4 dimensional Planck length and mass.

\subsection{Tidal Reissner-Nordstrom}

The ``Tidal" Reissner-Nordstrom metric \cite{dmptidalrn} comes from the showing that all the contraints known for $\xi_{\mu \nu}$ (symmetric, trace-free, and conservation equations) are satisfied by the stress energy tensor associated with the Einstein-Maxwell solutions of the standard 4 dimensional Einstein equations. By making the equivalence
\begin{equation}
\kappa^2 T_{\mu \nu} \leftrightarrow - \xi_{\mu \nu}
\end{equation}
and using the appropriate choice of constants, a solution to Eq. (\ref{eq:induced}) is found:  
\begin{equation}
ds^2 = -(1-\frac{2M}{r} + \frac{Q}{r^2})dt^2 + (1-\frac{2M}{r}+  \frac{Q}{r^2})^{-1} dr^2 +r^2 d\Omega^2
\label{eq:trnmetric}
\end{equation}
 Unlike the Reissner-Nordstrom solution, $Q$ here is allowed to be negative, strengthening the effect of gravity. Since this metric is taken to apply only in the strong field, as its asymptotic behavior is incorrect for non-trivial values of $Q$, tidal charge is weakly limited by studies of neutron stars \cite{trnconstraints} and perhaps potential observations of event horizon structure  \cite{NatureHorizon} which are still inconclusive. The TRN metric is not favored for large black holes according to \cite{wiseman}, but it is an interesting candidate for the strong field because the $\frac{Q}{r^2}$ term encodes a 5 dimensional potential that one would expect to see in a Randall-Sundrum braneworld. Studies \cite{whisker2005, mmreview} about the lensing effects of the TRN metric around Sgr A* have set $ Q= q\:4M^2$ where $-1<q<0$, as this allows for significant effects from the braneworld term. 

\section{\label{sec:bbhl} Black Hole Lensing in the Braneworld}

The greatest impetus for the study of relativistic images, images formed deeped within the strong deflection regime near the photon sphere of a black hole, is the ability to test GR against an alternative theory of gravity. Photons that reach us after probing close to a black hole are most likely to carry signs of gravity's true nature. There have been previous studies of strong deflection lensing around braneworld black holes \cite{whisker2005, eiroad, mmreview}. However, all use some variant of the analytical formalism developed by \cite{bozza2002, bozza2007}. In this section, we will calculate lensing observables in two scenarios: The SMBH at the center of the galaxy lensing a galactic source and a small primordial black hole at solar system scales lensing a galactic source. For both, we will use the numerical technique.

 In order to choose an appropriate metric for calculating relativistic image properties, we need to calculate the location of the photon sphere of a black hole. This is because relativistic images need to have a very large bending angle (at least $\pi$ for what \cite{eiroad} terms ``retro-lensing"). Such a large bending angle requires that the coordinate of closest approach for the null geodesic be very close to the photon sphere. Hence, knowledge of that scale tells us what regime we are dealing with and, therefore, which metric is appropriate. The photon sphere in braneworld metrics is given by setting  (Eq. (11) in \cite{VE2002}):

\begin{equation}
2 C(r) B(r) + r \frac{d C(r)}{dr} B(r) -r  \frac{d B(r)}{dr} C(r) =0
\end{equation}
This yields

\begin{eqnarray}
\nonumber  r_{ps}^{GT} &=& M+ \frac{3^{\frac{1}{3}} M^2}{(-5l^2M+ 3M^3+\sqrt{5}\sqrt{5l^4M^2+6l^2M^4})^{\frac{1}{3}}}+\\
&& \frac{(-5l^2M+ 3M^3+\sqrt{5}\sqrt{5l^4M^2+6l^2M^4})^{\frac{1}{3}}}{ 3^{\frac{1}{3}}}\\
r_{ps}^{MP} &=& \sqrt{2}\: r_H\\
r_{ps}^{TRN} &=& \frac{1}{2}(3M + \sqrt{9M^2+8Q}) \label{eq:trnps}
\end{eqnarray}
Note that these are all different than in the Scwarzschild metric in which $r_{ps}= 3M$. We now follow the procedure outlined in Section II but use the appropriate braneworld metric.

\subsection{Supermassive Black Hole}

When considering the black hole at Sgr A* as a lens, we use the values $D_L=8.3$ kpc and $M= 4.3 \times 10^6 M_{\odot}$ \cite{propSgrA}. We model the source as a point source at $D_S= 2 D_L$. Modelling the supermassive black hole (SMBH) at Sgr A$^{\ast}$ as a braneworld black hole can  be done with the Garriga-Tanaka metric, as the photon sphere is of order $10^7 $ km  which is much greater than the scale of the extra dimension $l$. However, since the correction to the Schwarzschild metric is of order $\frac{Ml^2}{r^3} \approx 10^{-31}$, the lensing corrections due to the Garriga-Tanaka metric is negligible as can be seen in Table \ref{ERcomparison}. Although some numerical studies indicate that such a large black hole is unlikely to have an exterior metric significantly different than Schwarzschild \cite{wiseman}, a case can be made for the Tidal Reissner-Nordstrom metric as the spacetime curvature near the surface of the SMBH is much greater than on the surface of the Earth and, therefore, is not excluded by our tests of Newtonian gravity on Earth. The curvature of spacetime is measured by the Kretschmann scalar (as the Ricci scalar is zero in the vacuum outside a black hole), which is for the Schwarzschild spacetime

\begin{equation}
K=R_{abcd}R^{abcd}= 48 \frac{m^2}{r^6}
\end{equation}

The ratio of the Kretschmann scalar calculated at the surface (photon sphere) of the SMBH to the Kretschmann scalar calculated at the surface of the Earth is:

\begin{equation}
\frac{K_{Sgr A^{\ast}}}{K_{\oplus}} \approx 2000
\end{equation}
This  implies that gravity near the black hole is very strong and braneworld gravity may show up at in strong gravitational fields while not being detected on Earth.  In addition, there are few constraints on positing a $\frac{1}{r^2}$ correction term to the gravitational potential that only exists in the strong field, so we can take the TRN metric as a specific example of a $\frac{1}{r^2}$ correction term to the metric. 
The TRN metric requires a choice as to the parametrization of the variable $Q$. In the literature \cite{eiroad, whisker2005, mmreview}, $Q$ is usually taken to have dimensions $Q \varpropto M^2$, so that the integral for the bending angle is dimensionless when rephrased in terms of Schwarzschild radii (See Eq. (\ref{bendingbw})). The dimensionless of the metric is desirable and therefore makes this parameterization of $Q$ a natural choice:

\begin{equation}
Q = q \: 4 M^2 
\label{eq:param2}
\end{equation}
with $|q|$ usually taking on values between $0$ and $1$. The weakness of this approach is that it makes the assumption that the strength of the bulk's ``backreaction" onto the brane is $ \varpropto M^2$. While this allows a neat result that eases calculations for relativistic observables and maximizes the braneworld effect by maximizing $Q$ in the case of large black holes, it is an ad hoc assumption. We will explore other parameterizations of the backreaction onto the bulk for other cases of the black hole. For the case of the SMBH, we will use this parameterization as it gives the maximal amount of tidal charge. The results of the location of the Einstein rings for different values of $q$ are contained in Table \ref{ERcomparison}. Since the first relativistic image appears very close to the position of the first relativistic Einstein ring, and higher-order relativistic images appear close to their corresponding Einstein ring, the position and separation of the first two relativistic Einstein rings is a good indication of the properties of image positions for relativistic images.

\begin{table*}[t]

\begin{tabular}{|c|c|c|c|c|}
\hline
\hline
  & Sch & Garriga-Tanaka & Tidal RN (q= -.1) & Tidal RN (q= -.5)\\
\hline
Einstein Ring  & 1.452 arcsec  &  1.452 arcsec & 1.452 arcsec & 1.452 arcsec   \\
1$^{\text{st}}$ Relativistic Einstein Ring & 26.57 $\mu$ arcsec & 26.57 $\mu$ arcsec & 28.22 $\mu$ arcsec & 33.38 $\mu$ arcsec   \\
2$^{\text{nd}}$ Relativistic Einstein Ring & 26.54 $\mu$ arcsec  & 26.54 $\mu$ arcsec & 28.19 $\mu$ arcsec &  33.37 $\mu$ arcsec  \\
\hline
\hline

\end{tabular}

\caption{\label{ERcomparison} This table calculates the properties of Einstein rings formed by a source directly aligned with Sgr A* ($M= 4.3 \times 10^6 M_{\odot}$ and distance $D_L = 8.3$ kpc) and with a source distance of $2D_L$.  This table contains the angular positions of the primary Einstein ring and two relativistic Einstein rings formed by the source. The location of second relativistic Einstein ring is very close to the photon sphere and the change in its location in different metrics is a reflection of the different structure of the photon sphere in the braneworld scenario. Note the identical results for the Garriga-Tanaka and Schwarzschild metrics. Also note the difference in the gap between the first and second relativistic Einstein ring in each geometry.}
\end{table*}

The effect of braneworld metric on image magnification can be seen in Table \ref{relimgcomp1}. Use of the TRN metric slightly demagnifies the images of a source almost directly directly behind ($\beta=1.45 \mu$as) at the distance $D_S =2 D_L$. For the higher value of tidal charge, the images are even more demagnified.   Repeating this calculation for multiple source positions ranging from 1 as to 1 $\mu$ as, the total magnification changes, but the ratio between the magnification in a Schwarzschild spacetime and a TRN spacetime for a specific value of $q$ remains the same.  For a SMBH, the GT metric will not produce any deviation from the standard GR results because the non-Schwarzschild term in the GT metric scales as $(\frac{l}{m})^2$. However, the TRN metric displays theoretically differentiable results, a result that encourages further study. The images in the TRN results are fainter than the corresponding images in the Schwarzschild metric for this particular case, but this is not at all a general characteristic of braneworld strong field images. This will be discussed further at the end of this section.

\begin{table*}[t]

\begin{tabular}{|c|c|c|c|}
\hline
\hline
  & Sch & TRN (q= -.1) &  TRN (q= -.5)  \\
\hline
Outer Rel. Image (opposite side of source)  & $-5.94 \times 10^{-12}$ &  $-5.54 \times 10^{-12}$ & $-4.48 \times 10^{-12}$    \\
Inner Rel. Image (opp. side of source) & $-1.10 \times 10^{-14}$  &  $-7.85 \times 10^{-15}$ & $-3.97 \times 10^{-15}$    \\
Inner Rel. Image (side of source) & $1.10 \times 10^{-14}$  & $ 7.85 \times 10^{-15}$ & $3.97 \times 10^{-15}$ \\
Outer Rel. Image (side of source) &  $5.94 \times 10^{-12}$ & $5.54 \times 10^{-12}$ & $4.48 \times 10^{-12}$  \\
\hline
\hline

\end{tabular}

\caption{\label{relimgcomp1} This table contains the magnifications of relativistic images formed by a source at twice the distance of Sgr A* (see caption of Table \ref{ERcomparison}) and at an angular position of $\beta=1.45$ $\mu$as or $10^{-3}$ times the Einstein angle in the Schwarzschild geometry for a directly aligned source. These properties are calculated using several different metrics. The values for the GT metric are indistinguishable from the Schwarzschild metric and, therefore, have not been shown.}

\end{table*}

Since we have chosen $Q$ in Eq. (\ref{eq:trnmetric}) to be parameterized by $q \varpropto \frac{Q}{4m^2}$, the explicit form of Eq. (\ref{bending}) for the TRN metric when using the equality in Eq. (\ref{r2mx}) is:
\begin{eqnarray}
\nonumber \alpha_{TRN}(x_0)&= &  \int_{x_0}^{\infty} \frac{2}{\sqrt{\frac{x}{x_0}(1-\frac{1}{x_0}+\frac{q}{x_0^2})-(1-\frac{1}{x} + \frac{q}{x^2})}} \\
& \times & \frac{dx}{x} - \pi
\label{bendingbw}
\end{eqnarray}
Since there is no dependency on mass for the bending angle in either the TRN or Schwarzschild ($q=0$) scenarios, there is no dependency on mass in the ratio of the magnifications (Eq. (\ref{magnification})) either. This is seen explicitly by considering Eq. (\ref{Mu}). As demonstrated by \cite{VE2000}, the position of relativistic images are very insensitive to source positions. Hence, the $x$ coordinate position of the $n^{th}$ relativistic image is very close to the value of the $x$ coordinate that yields a bending angle of $2 \pi n$ and the angular position of this image, $\theta$, scales with $M$. If we hold $\beta$ to be constant,  the term $\frac{\theta}{\beta}$ in  Eq. (\ref{magnification}) will scale as $M$. When we consider $\frac{d\theta}{d\beta}$, it is proportional to $m \frac{dx}{d\beta}$.  $\frac{dx}{d\beta}$ is constant in $\beta$ (see graphs of image position against source position in \cite{VE2000}) and therefore the entire term scales as $M$. So the magnification of a relativistic image for a constant source position scales as $M^2$. We can repeat the same argument for the bending angle in the braneworld scenario because Eq. (\ref{bendingbw}) does not contain any dependency on $M$. Magnification would then have a scaling of $M^2$ as well. Hence, further studies of this form of the TRN metric will not yield interesting results, as the difference between a Schwarzschild and a TRN lens does not change with the mass of the lens. In the next section, we will change the parameterization of the tidal charge for smaller black holes and anaylze the difference this makes for lensing properties.

\subsection{Primordial Black Holes}

One explanation for dark matter is primordial black holes (PBHs) that can form from a variety of mechanisms \cite{pbhhawking, phasereview}. Gould \cite{femtolensing} proposes femtolensing of gamma ray bursts, which are usually extra-galactical \cite{gammarates}, by compact dark matter. If PBHs make up a large fraction of dark matter in the universe, there is a chance of observing a gamma-ray burst with an interference pattern characteristic of lensing by a small black hole. An interference pattern resulting from such a lensing  could potentially be observed by the ongoing Fermi Gamma-ray Space Telescope. This idea is extended to the braneworld black hole scenario by \cite{petterskeeton}, who examine the effects of a GT metric of the interference pattern. A problem with analyzing a primordial black hole of the size considered by \cite{petterskeeton}, $10^{-18} M_{\odot}$ is that relativistic images will pass in a very highly curved area near the black hole. The radius of curvature will be of order $10^{-14}$ m, which is smaller than the wavelength of all electromagnetic radiation except for hard gamma rays, rendering the treatment unsuitable for the geometric optics approximation applied in this paper \cite{petters, schneider}. Therefore, we will consider a black hole of mass $10^{-14} M_{\odot}$. In the Schwarzschild metric, the horizon size will be about $3 \times10^{-11}$ m and we can consider it to have some validity for even soft gamma rays. In the TRN spacetime, the horizon size will be even larger (about $10^{-6}$ m for $q=-.5$), making our results applicable for the entire ultraviolet spectrum.  Since the point of closest approach is smaller than the scale of the extra dimension ($r << l$), I will ignore the Garriga-Tanaka metric and focus on the TRN and Myers-Perry metrics. 

While we can use the same parameterization as in Eq. (\ref{eq:param2}), as explained before, the results would be qualitatively similar. Also, it would be interesting to explore a scenario in which $Q$ does not scale strictly as $M^2$. In the paper introducing tidal charge, \cite{dmptidalrn} considers a scenario in which tidal charge is fixed and does not depend on the mass of the gravitating body. For such a source of tidal charge, they cite an upper bound for $Q$ of 

\begin{equation}
Q << 2  M_{\odot} R_{\odot}
\label{eq:tidalbound}
\end{equation}
because higher values for $Q$ will violate bounds on solar system tests if this metric is applied to our sun. This bound considers the possibility that $Q$ is a fixed feature of geometry and does not scale with mass (this bound does not apply if the TRN is proposed to only be the strong field limit of the metric). If we introduce this idea into our parameterization, the behavior of relativistic images in the braneworld scenario diverges from the behavior of Schwarzschild black holes. 

Now, I would like to consider a hybrid scenario. Let us say that we want $Q \varpropto M$ instead of $M^2$, we would then need to introduce a mass scale in order to keep the metric and Eq. (\ref{bendingbw}) dimensionless. This will be a hybrid of tidal charge: part of it will come from mass getting reflected back at a linear rate and part will come from a fixed feature of the geometry. If Eq. (\ref{eq:tidalbound}) is to be satisfied, we must have $M_{scale} << M_{\odot}$ so I set it as

\begin{equation}
M_{scale} =10^{-3} M_{\odot} \approx 1 m
\label{massscale}
\end{equation}
then we can introduce the parameterization:

\begin{equation}
Q = q \:  \: 2M \: M_{scale}
\label{eq:massscale2}
\end{equation}
This makes the bending angle integral:

\begin{eqnarray}
\nonumber \alpha_{TRN}(x_0) &  = & \int_{x_0}^{\infty}  \frac{2}{\sqrt{\frac{x}{x_0}(1-\frac{1}{x_0}+\frac{q M_{scale}}{2 M x_0^2})-(1-\frac{1}{x} + \frac{q M_{scale}}{2 M x^2})}}\\
& \times & \frac{dx}{x} - \pi 
\label{bendingbw2}
\end{eqnarray}
It is evident from Eq. (\ref{bendingbw2}) that braneworld effects are inversely proportional to the lens mass. We will consider sources at solar system scales. We will model our lens as a primordial black hole at 1 AU and our source at 1 Megaparsec. The source angular position $\beta$ is set as one millionth of the Einstein angle for this configuration (in this case, $10^{-12}$ as). In Table \ref{primordialcomp} we compare the difference between relativistic images in the Schwarzschild and braneworld spacetimes. The large amount of tidal charge relative to the mass of the black hole magnifies images near braneworld black holes by several orders of magnitude compared to images near their Schwarzschild counterparts.

\begin{table*}[t]
\begin{tabular}{|c|c|c|c|c|}
\hline
\hline
  & Sch &  TRN (q= -.1) &   TRN (q= -.5)& Myers-Perry \\
\hline
Outer Rel. Image (opposite side of source)  & $-1.66 \times 10^{-29}$ &  $-1.95 \times 10^{-21}$ & $-9.74 \times 10^{-21}$ &   $-8.97 \times 10^{-25}$ \\
Inner Rel. Image (opp. side of source) &  $-3.08 \times 10^{-32}$  &  $-2.69 \times 10^{-25}$ & $-1.35 \times 10^{-24}$    & $-1.24 \times 10^{-28}$\\
Inner Rel. Image (side of source) &  $3.08 \times 10^{-32}$  &  $2.69 \times 10^{-25}$ & $1.35 \times 10^{-24}$ &$1.24 \times 10^{-28}$\\
Outer Rel. Image (side of source) &  $1.66 \times 10^{-29}$ &  $1.95 \times 10^{-21}$  & $9.74 \times 10^{-21}$ &  $8.97 \times 10^{-25}$ \\
\hline
\hline

\end{tabular}

\caption{\label{primordialcomp} This table calculates the magnification of the outer two relativistic images on each side of the optic axis for a primordial black hole at 1 AU lensing a distant source at 1 Megaparsec, with $\beta= 10^{-12} \mu$ as}
\end{table*}

\subsection{Magnification in the Braneworld}

In the previous sections, we have shown that for the TRN metric, relativistic images in the braneworld are slightly fainter for relativistic images around supermassive black holes and are greatly enhanced for primordial black holes. The distinction between the results in the case of primordial and supermassive black holes benefits from looking at the two components in Eq. (\ref{Mu}). The two component magnifications are termed tangential magnification ($\mu_t$) and radial magnification ($\mu_r$) and are defined as

\begin{eqnarray}
\mu_t & \equiv & (\frac{ \text{Sin } \beta }{  \text{Sin } \theta})^{-1}\\
\mu_r & \equiv & (\frac{d\beta }{d\theta})^{-1}
\end{eqnarray}
\cite{bozza2002} approximates both of these quantities and puts them in analytical form:

\begin{eqnarray}
\mu_t & = & (\frac{\beta}{\theta_n^0})^{-1} \label{eq:mut}\\
\mu_r & = & (1+\frac{\bar{a} D_L D_{LS}}{u_m e_n D_S})^{-1}
\label{eq:mur}
\end{eqnarray}

From Eq. (\ref{eq:mut}), the tangential magnification of a relativistic image is directly proportional to the image position of the relativistic image. This will always be larger when using a braneworld metric because of the strengthened gravity in the braneworld metric. Since for primordial black holes, $Q$ can be very large relative to the mass of the black hole, the photon sphere is approximated from Eq. (\ref{eq:trnps}) as $r_{ps}^{TRN} \approx \sqrt{8Q}$ when $Q \gg M$. So, the tangential magnification is always larger in the braneworld scenario, sometimes considerably greater.

The variation in behavior of braneworld magnifications as mass changes comes from the radial magnification. From Eq. (\ref{eq:mur}), it is apparent that for small radial magnifications (which is always the case for relativistic images), radial magnifications are proportional to:

\begin{equation}
\mu_r \varpropto \frac{u_m e_n}{\bar{a}}
\label{eq:radialterm}
\end{equation}
This relation can be physically motivated- $u_m$ is a measure of the photon sphere's size. This is because the coordinate of relativistic image is close to the location of the photon sphere. Magnification of the relativistic image is $\varpropto \frac{dx}{d \theta}$, so when $x$ is rescaled to reflect a larger photon sphere, the derivative becomes larger as well. The term $e_n$ defines how quickly images decay towards towards the photon sphere for the n$^{th}$ image. If $e_n$ is greater, the image will shift more with a change in source position and therefore,the radial magnification will be brighter. The image brightness is also inversely proportional to $\bar{a}$.

\begin{figure}[h]
 \begin{center}
 \includegraphics[width=0.5 \textwidth]{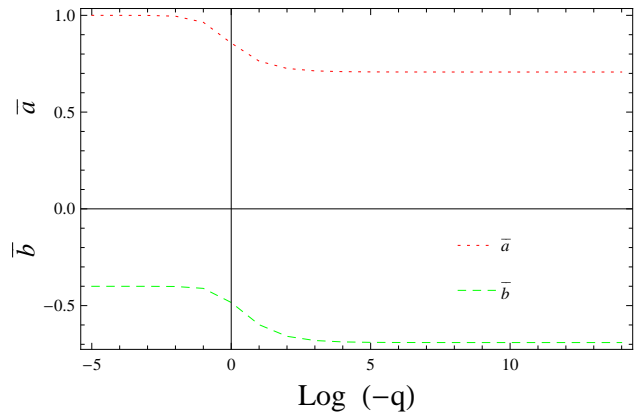}
 \caption{\label{fig:sfo}  The behavior of strong field quantities as the behavior becomes 5 dimensional. The dotted line is the quantity $\bar{a}$ and the dashed quantity is $\bar{b}$. From left to right, the quantity $q$ becomes a progressively larger negative number.}
 \end{center}
 \end{figure}

It is simple to compute the behavior of each of these quantities in a braneworld scenario. In Fig. \ref{fig:sfo}, we show the behavior of quantities $\bar{a}$ and $\bar{b}$ as $q$ gets larger. The behavior of $u_m$ for large $|q|$ is given by Eq. (\ref{eq:trnps}).

As can be seen from Fig. \ref{fig:sfo}, the quantity $\bar{a}$ is equal to 1 for a Schwarzschild metric and becomes smaller as $-q$ becomes large and asymtotes to $\frac{1}{\sqrt{2}}$. The quantity $\bar{b} = -.400$ in a Schwarzschild spacetime, but asymptotes to $-.691$ for a 5 dimensional spacetime. Therefore, $e_n$ (for $n=1$) varies from $.00125$ in the Schwarzschild limit to $.0000521$ in the $q \rightarrow -\infty$ limit. When tidal charge is weaker (supermassive black hole), $u_m$ is not much larger than the $q=0$ value, the smaller $e_n$ term dominates and the relativistic image is demagnified. When there is a great deal of tidal charge ($Q \gg M$), the great increase in $u_m$ makes up for the $e_n$ term and the relativistic image is brighter compared to its Schwarzschild counterpart.

For all black holes, relativistic images will be at a larger image position when using a braneworld metric, because the photon sphere of a braneworld metric is larger than its Schwarzschild counterpart. Hence, the tangential magnification is always higher using a braneworld. Radial magnification, as we have shown, depends on the value of the tidal charge. For a given source and lens position, the radial magnification is lower bounded, even as $|q|$ gets very large, but the tangential mangification is not upper bounded. For a large enough tidal charge, magnification will be enhanced for a braneworld image.

\section{\label{sec:ana} Applicability of the Analytical Method}

\begin{figure}[h]
 \begin{center}
 \includegraphics[width=0.4 \textwidth]{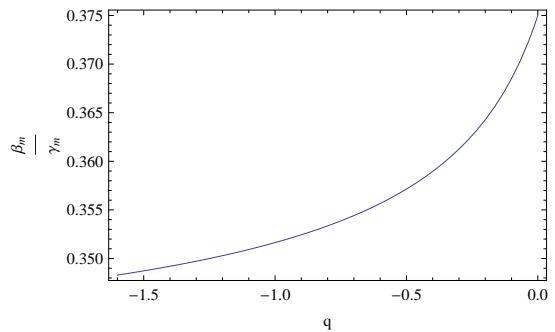}
 \caption{The behavior of $\frac{\beta_m}{\gamma_m}$  as -$q$ becomes larger. The decrease in its value shows that the logarithmic approximation for the bending angle becomes worse as $q$ gets smaller.}
 \label{fig:ratio}
 \end{center}
 \end{figure}

In Sec. \ref{sec:sfl}, constructing a logarithmic approximation for the bending angle depends on the relationship in Eq. (\ref{eq:relation}) between the impact parameter and the minimum radial coordinate. When expanding $R(r_0, u)=0$ in orders of $\epsilon$ and $\delta$, Eq. (\ref{eq:firstorder}) is obtained by truncating at the first non-zero term in both $r$ and $u$. However, the relationship between $\delta$ and $\epsilon$ is not being considered at an infinitesimal distance from the photon sphere. For example, in a Schwarzschild geometry, the first relativistic Einstein ring occurs with approximately $r_0 \approx 3.09 M$ \cite{VE2000}, which corresponds to $\delta= .03$. This corresponds to $\epsilon= -.0013$. To be able to approximate the bending angle with a logarithmic term, we must be able to neglect the higher order terms that do not appear in Eqs. (\ref{eq:firstorder}) and (\ref{eq:relation}). While \cite{bozza2002, bozza2007} show that this first order treatment is adequate for typical spherically symmetric metrics such as the Schwarzchild, Reissner-Nordstrom (RN), and Janis-Newman-Winicour  (JNW) metrics. However, some signs of this approximation breaking down is seen for RN and JNW metrics when they become near-extremal \cite{bozza2002}. 
Our goal in this section is to perform a general analysis of the higher order terms in the general case and then apply it to show the validity of the logarithmic approximation in the case of the TRN metric. To analyze the general case, we will expand out Eq. (\ref{eq:firstorder}) for one more order in both $\epsilon$ and $\delta$, which yields 
\begin{equation}
\beta_m \delta^2 + \gamma_m \delta^3= 2 u_m^2 \epsilon + u_m^2 \epsilon^2
\label{eq:higherrelation}
\end{equation}
where
\begin{equation}
\gamma_m= \frac{1}{6} \frac{\partial^3 R}{\partial r^3}(r_m, u_m) r_m^3
\label{eq:higherorder}
\end{equation} 
First, consider the right side of Eq. (\ref{eq:higherrelation}). The higher order term $ \frac{\partial^3 R}{\partial u^3}(r_m, u_m)$ is 0 from the definition of $R(r_0, u)$.We can neglect the $\epsilon^2$ term for the following reason. The ratio of the $\epsilon$ terms is known- the ratio of the first-order term to the second order term is $\frac{2}{\epsilon}$. This remains fixed regardless of the geometry because the $u^2$ term in $R(r_0, u)$ is independent of the metric. In the Schwarzschild case, $\epsilon$ is much smaller than $\delta$ and, therefore, higher order terms in $\epsilon$ are less significant than higher-order terms in $\delta$. This holds true in most spacetimes and can be checked using Eq. (\ref{eq:relation}).
Examining the left side of Eq.(\ref{eq:higherrelation}), the ratio between the second order and third order terms in $\delta$ is
\begin{equation}
\frac{\beta_m}{\gamma_m \delta}
\end{equation}
The greater this quantity is, the less significant the third order term is. Since $\beta_m$ and $\gamma_m$ are functions of the metric, the ratio of thes two terms depends on the underlying spacetime. Since the validity of this approximation scheme is known for the Schwarzschild metric, it is useful to compare the ratio $\frac{\beta_m}{\gamma_m}$ in different spacetimes to its ratio in the Schwarzschild spacetime. For a Schwarzschild metric:
\begin{equation}
\frac{\beta_m}{\gamma_m}= \frac{3}{8}
\label{eq:ratiosch}
\end{equation}
This result explains why the approximation scheme in Sec. \ref{sec:sfl} only holds up for points of closest approach that are close to the photon sphere (small $\delta$). For large $\delta$, the higher-order terms remain significant. However, as demonstrated in \cite{bozza2002, bozza2007}, it can be considered valid for the domain of relativistic images. Hence, we know that for the ratio in Eq. (\ref{eq:ratiosch}), we can consider the bending angle approximation to be valid. If, for an alternate spacetime, this ratio becomes bigger, the approximation will be better. If the ratio becomes smaller, the approximation will fare worse.
For the TRN metric, the ratio $\frac{\beta_m}{\gamma_m}$ depends on the parameter $q$. As -$q$ gets larger, the ratio gets smaller as well and the approximation gets worse as well. The analytical expression for both $\beta_m$ and $\gamma_m$ can be obtained easily. Figure \ref{fig:ratio} displays the relationship between $q$ and $\frac{\beta_m}{\gamma_m}$  
Fig. \ref{fig:ratio} shows that as $q$ gets smaller, $\frac{\beta_m}{\gamma_m}$  drops from its starting value of $\frac{3}{8}$ in a Schwarzschild metric. For the parameterization in Eq. (\ref{eq:massscale2}), the amount of tidal charge can be very large. We can evaluate the performance of the approximation in the limit of large tidal charge:
\begin{equation}
\lim_{q \rightarrow -\infty}\frac{\beta_m}{\gamma_m}= \frac{1}{3}
\label{eq:limit}
\end{equation}
This shows that for any amount of negative tidal charge, the approximation will be worse than in the Schwarzschild case, but only marginally so. This is borne out by an actual comparison of results obtained by both methods.  As $\delta$ gets larger, the validity of the approximation falls off quicker in the TRN spacetime. As $q \rightarrow -\infty$, the TRN metric becomes a 4 dimensional metric, like the Myers-Perry metric, for which strong deflection limit lensing has been examined \cite{mmreview, eiroad}. The result in Eq. (\ref{eq:limit}) shows that the strong deflection limit is accurate for the TRN metric and for the Myers-Perry metric.  For any spherically symmetric metric, one test for this approximation scheme can be the examination of $\frac{\beta_m}{\gamma_m}$ in that spacetime.

 \section{\label{sec:conc} Conclusion}

In this paper, we have shown various properties of relativistic images and how the qualitative features of black hole lensing for a braneworld black hole depend on the size of the black hole. We have introduced a new parameterization of ``tidal charge" for the Tidal RN metric and have thereby uncovered a richer phenomenology of braneworld black hole structure in the strong deflection limit. We show that the smaller the black hole is, the greater the effect of the braneworld scenario is on the relativistic image observables. The lensing properties of small black holes are especially relevant with the start of the LHC, which has the possibility of producing TeV-scale black holes in the braneworld scenario. While the lensing properties of such a small black hole cannot be calculated using the geometric optics approach employed in this paper, there is good reason to suspect that lensing behavior in the braneworld will be dramatically different than Schwarzschild behavior. We have shown that relativistic images are greatly enhanced by the braneworld geometry, but it remains an open question whether there are observational consequences to the image enhancement. 

We also confirm that the analytical method serves as an accurate probe of relativistic images, within a reasonable margin of error. At this stage, when the observational possibilites for relativistic images are more about the observation of the images than cataloguing the exact positions and magnitudes, a qualitative description of relativistic images is enough to explore this topic further. An advantage of the analytical approximation is that it allows for the calculation of certain observables more easily then using the numerical method. For example, the sum of the magnitude of all relativistic images after the first (which should appear unresolved) is easily accomplished using the analyitical methodology. The analytical formulation also can be easily applied to extended sources \cite{eiroad}, which is more difficult with the numerical methodology. However, use of a numerical integral is necessary in the astrophysically relevant situation \cite{bozzamancini2009, bozzas2, bozzas2revised} of secondary images of S stars oribiting Sgr A$^{\ast}$ that undergo a bending angle in the middle of the region $0<\alpha<\pi$. Here, both the weak and strong deflection approximations fail and evaluating integrals numerically is required for accurate calculation of primary and secondary image properties. Hence, both methodologies will be relevant for the calculation of potentially observable consequences of relativistic images.

It has been less than 10 years since the possibility of observing relativistic images and other strong field images has been considered in \cite{VE2000}, and these images represent a promising source for knowledge about gravity in its full strength. Observation of Sgr A$^{\ast}$ is a major area of research, and a theoretically interesting lensing observable can pave the way for attempts to use high-resolution VLBI arrays to find these elusive images. Upcoming instruments such as GRAVITY and MICADO at the European Extremely Large Telescope (E-ELT) also have the potential to make observations of strong deflection limit lensing near Sgr A$^{\ast}$. Observation of a single relativistic image has the potential to answer deeply fundamental questions such as whether there is an extra dimension, scalar charge, or even black holes. The immense theoretical potential for relativistic observables in the vicinity of Sgr A$^{\ast}$ should motivate effort to continue to refine theoretical models and lensing scenarios.

\begin{acknowledgments}
I would like to thank R.K. Sheth for his unwavering support at all stages of research and writing. I thank K.S. Virbhadra for suggesting some aspects of this study, and help at the initial stages of this work. Thanks are due to J. Guzik, J. Hyde, J. Khoury and C. Dinerman for helpful discussions and to A. Congdon for reviewing the manuscript. I would also like to thank G. Cwilich and Yeshiva University for hosting me during research for this paper. 
\end{acknowledgments}

\bibliographystyle{h-physrev}

\end{document}